\pdfoutput=1

\documentclass[11pt,twoside,a4paper,cmspaper,final,collab]{cms-tdr}
\def\svnVersion{a8d4d55-D}\def\svnDate{2019/08/20}\def\cmsCernNoTag{CERN-EP-2018-268}\def\cmsCernDate{\today}\def\cmsMessage{"Published in Physical Review C as \href{http://dx.doi.org/10.1103/PhysRevC.100.024902}{\doi{10.1103/PhysRevC.100.024902}.}"}

\begin{document}\cmsNoteHeader{HIN-14-014}

\hyphenation{had-ron-i-za-tion}
\hyphenation{cal-or-i-me-ter}
\hyphenation{de-vices}
\RCS$HeadURL$
\RCS$Id$

\newlength\cmsFigWidth
\newlength\cmsTabSkip\setlength{\cmsTabSkip}{1ex}
\ifthenelse{\boolean{cms@external}}{\setlength\cmsFigWidth{0.85\columnwidth}}{\setlength\cmsFigWidth{0.4\textwidth}}
\ifthenelse{\boolean{cms@external}}{\providecommand{\cmsLeft}{top}}{\providecommand{\cmsLeft}{left}}
\ifthenelse{\boolean{cms@external}}{\providecommand{\cmsRight}{bottom}}{\providecommand{\cmsRight}{right}}

\newcommand{\GeVfmcube}{\ensuremath{\GeV/\text{fm}^3}\xspace}
\newcommand{\pPb}{\ensuremath{{\Pp}\text{Pb}}\xspace}
\newcommand{\Pbp}{\ensuremath{\text{Pb}{\Pp}}\xspace}
\newcommand{\pU}{\ensuremath{{\Pp}\text{U}}\xspace}
\newcommand{\pAu}{\ensuremath{{\Pp}\text{Au}}\xspace}
\newcommand{\pA}{\ensuremath{{\Pp}\text{A}}\xspace}

\newcommand{\dA}{\ensuremath{{\PQd}\text{A}}\xspace}
\newcommand{\dAu}{\ensuremath{{\PQd}\text{Au}}\xspace}

\cmsNoteHeader{HIN-14-014}
\title{Centrality and pseudorapidity dependence of the transverse energy density in \texorpdfstring{\pPb collisions at
$\sqrtsNN=5.02\TeV$}{pPb collisions at sqrt(s[NN]) = 5.02 TeV}}

\date{\today}

\abstract{
The almost hermetic coverage of the CMS detector is used to measure the
distribution of transverse energy, \ET, over 13.2 units of pseudorapidity, $\eta$,  for \pPb collisions at a center-of-mass energy per nucleon pair of
$\sqrtsNN=5.02\TeV$.
The huge angular acceptance exploits the fact
that the CASTOR calorimeter at $-6.6<\eta<-5.2$ is effectively present
on both sides of the colliding system because of a
switch in the proton-going  and lead-going beam directions.
This wide acceptance
enables the study of correlations between well-separated
angular regions and
makes the measurement a particularly
powerful  test  of  event generators.
For minimum bias \pPb collisions
the maximum value of
$\rd\ET/\rd\eta$  is  22\GeV,   which implies an \ET
per participant nucleon pair  comparable to that of peripheral PbPb collisions at $\sqrtsNN=2.76\TeV$.
 The increase of
 $\rd\ET/\rd\eta$
 with centrality is much stronger  for the lead-going side than for the proton-going side.
 The $\eta$ dependence of  $\rd\ET/\rd\eta$ is sensitive to the $\eta$ range in which the centrality variable is defined.
 Several modern generators are compared to these results but none is able
 to capture all aspects of the
 $\eta$
 and centrality dependence of the data and the correlations observed between different $\eta$ regions.
 }

\hypersetup{
pdfauthor={CMS Collaboration},
pdftitle={Centrality and pseudorapidity dependence of transverse energy density in pPb collisions at
sqrt(s[NN])=5.02 TeV},
pdfsubject={CMS},
pdfkeywords={CMS, heavy ion physics}}

\maketitle

\section{Introduction}

In a heavy ion or proton nucleus collision the total transverse energy, \ET, is a measure of the energy liberated by the deceleration, or ``stopping power" of the colliding nucleons while $\rd\ET/\rd y$ measures the
total energy carried by the system of particles or medium, produced in the collision, which is moving with longitudinal rapidity $y$~\cite{Busza:2018rrf}.
In heavy ion collisions the energy density, $\epsilon_\text{BJ}$, of this medium at proper time
$\tau_0$
shortly after the impact of the two nuclei
can be estimated using the Bjorken formula
\begin{linenomath*}
   \begin{equation}
   \epsilon_\text{BJ}=\frac{\rd\ET}{\rd y}\frac{1}{\tau_0 A_\perp},
   \label{Eqn:EnergyDensity}
  \end{equation}
\end{linenomath*}
where
$ A_\perp$ is the nuclear transverse area, \ie, the initial size of the medium~\cite{PhysRevD.27.140}.
The time $\tau_0$ at which it is first appropriate to speak about an energy density is a model
assumption.
Some collaborations have chosen to report the product of energy density and  proper time $ \epsilon_\text{BJ} \tau_0$ \cite{PhysRevD.27.140,Adam:2016thv} while others have used  $\tau_0=1\unit{fm}/c$ as a reference value \cite{Chatrchyan:2012mb,Margetis:1994tt}.

For the   top 5\% most central lead-lead
collisions at
$\sqrtsNN=2.76\TeV$,
this formula gives  energy densities up to  14\GeVfmcube
at a time
$\tau_0=1\unit{fm}/c$~\cite{Chatrchyan:2012mb}.
This value is above the expected threshold of
$\epsilon>1\GeVfmcube$
for the production of a quark-gluon plasma estimated from
quantum chromodynamics, QCD,
calculations
performed on a lattice~\cite{Karsch:2001cy}.
Collective phenomena such as azimuthal flow and strangeness enhancement have been observed in
 proton-lead (\pPb)~\cite{Khachatryan:2015waa,Aad:2013fja,Acharya:2017tfn}
and even high-multiplicity proton-proton (\Pp\Pp) collisions~\cite{Khachatryan:2010gv,Khachatryan:2015lva,Khachatryan:2016txc,Aaboud:2017acw,Adam:2015vsf,Aad:2015gqa}.
Given such evidence of collective motion  and strangeness enhancement in small systems,
it is relevant to study the energy
densities  achieved
in \pPb collisions to see if a quark-gluon plasma could be formed in \pPb collisions.

The \ET spectra
in proton-nucleus,  \pA and deuteron-nucleus, \dA, collisions have been measured at
 center-of-mass energies ranging  from
$\sqrtsNN=5.5$ to 200\GeV
with nuclei
ranging from deuterium (atomic number $A=2$)  to uranium (U, $A=238$)~\cite{Abbott:1988kk,Barrette:1995hq,Akesson:1992uv,Adare:2015bua}.
At  $\sqrtsNN=5.5\GeV$,
only a weak correlation is observed between the total  \ET and the charged-particle  multiplicity in the forward region~\cite{Barrette:1995hq}.
At  $\sqrtsNN=5.5$, 20,  and 30\GeV, the mean pseudorapidity
$\eta$
moves backward, \ie, in the ion-going direction,  and the pseudorapidity width of the
$\rd\ET/\rd\eta$  distribution decreases as the
 total $\ET$ in the event increases~\cite{Abbott:1988kk,Barrette:1995hq,Akesson:1992uv}.

In this paper, we report
$\rd\ET/\rd\eta$ distributions measured in \pPb collisions at  $\sqrtsNN=5.02\TeV$ by the CMS experiment at the CERN LHC.  This beam energy is 25 times larger than that
for the previous highest energy measurements at RHIC~\cite{Adler:2013aqf}. The analysis combines
measurements from both \pPb and \Pbp data taking to
 cover
 13.2 units of $\eta$, \ie, $\abs{\eta} < 6.6$ in the laboratory frame.
 Since the energy per nucleon of the  proton beam is higher than that of the lead one, the nucleon-nucleon center-of-mass is at a pseudorapidity of $\eta_\text{lab}=0.465$  in the laboratory frame of reference.
 For symmetric heavy
ion collisions, the shape of $\rd\ET/\rd\eta$ \vs $\eta$ has only a weak dependence  on the $\eta$ region,
 which is used to classify the centrality of the  events~\cite{Chatrchyan:2012mb}.    To test if this is the case for the much smaller system created in \pPb collisions,  events are classified according to the \ET or charged-particle multiplicity in several  different $\eta$ regions, and the $\rd\ET/\rd\eta$ distributions produced by the different classification procedures are compared to each other.

  The comparison of these collider data with modern event generator calculations  is a significant motivation
for this work. The data presented here reach into the forward region that is crucial for
 understanding  the development of cosmic ray air showers.
A significant uncertainty in cosmic ray physics arises from the simulation of very high energy
hadron-air collisions~\cite{Ulrich:2010rg}.
This uncertainty has an important effect on the modeling of air showers and the energy calibration of modern cosmic ray observatories.
For a proper description of the development of cosmic ray air showers it is crucial to understand the rapidity region within four units of the rapidity of the incoming proton or nucleus~\cite{Kheyn:2012my}.
The data are compared in detail to calculations from three event generators: \textsc{hijing}  v2.1,
\textsc{epos-lhc} and \textsc{qgsjet ii}-04~\cite{Wang:1991hta,Pierog:2013ria,Ostapchenko:2010vb}.

\section{The CMS apparatus}
\label{Sec:Apparatus}

The central feature of the CMS apparatus is a superconducting solenoid of 6\unit{m}
internal diameter, providing a magnetic field of
3.8\unit{T}.
Within the solenoid volume are a silicon pixel and strip tracker, a lead tungstate crystal electromagnetic calorimeter (ECAL), and a brass and scintillator hadron calorimeter (HCAL), each composed of a barrel and two endcap sections. Muons are measured in gas-ionization detectors embedded in the steel flux-return yoke outside the solenoid.  The silicon detectors provide tracking in the
region $\abs{\eta} < 2.5$,  ECAL and HCAL cover the pseudorapidity interval  $\abs{\eta} < 3.0$ while the muon system covers the region  $\abs{\eta} < 2.4$.
In the forward region, the hadron forward (HF) calorimeters cover the region
$3.0<\abs{\eta} < 5.2$.

Each HF calorimeter consists of 432 readout towers, containing long and short quartz fibers running parallel to the beam.
By reading out the two sets of fibers separately, it is possible to distinguish showers generated by electrons and photons from those generated by hadrons. Very forward angles are covered at one end of CMS ($-6.6 < \eta < -5.2$) by the CASTOR calorimeter, and at both ends ($\abs{\eta} > 8.3$) by the zero-degree calorimeters (ZDCs). Both CASTOR and the ZDCs
consist of  quartz plates or fibers embedded in tungsten absorbers.
 They are segmented longitudinally to allow the separation of electromagnetic and hadronic components of the showers produced by incoming particles.
A more detailed description of the CMS detector, together with a definition of the coordinate system used and the relevant kinematic variables, can be found in Ref.~\cite{Chatrchyan:2008zzk}.

Analysis in the midrapidity region is based upon objects produced by  the
CMS particle-flow algorithm~\cite{Sirunyan:2017ulk}, which reconstructs and identifies each individual particle-flow candidate with an optimized combination of information from the various elements of the CMS detector.
The energy of photons is directly obtained from the ECAL measurement, corrected for  the effects of
the zero-suppression algorithm. The zero-suppression algorithm both speeds up the  readout and reduces the volume of data that must be recorded. The energy of electrons is determined from a combination of the electron momentum at the primary interaction vertex, as determined by the tracker, the energy of the corresponding ECAL cluster, and the energy sum of all bremsstrahlung photons
compatible with originating from the electron track. The energy of muons is obtained from the curvature of the corresponding track, reconstructed using information from both tracker and muon stations.
For $\abs{\eta}<2.5$ the
 energy of charged hadrons is determined from a combination of their momentum measured in the tracker and the matching of ECAL and HCAL energy deposits. These energy deposits are corrected for the effects of
the zero-suppression algorithm and the response function of the calorimeters to hadronic showers. Finally, the energy of neutral hadrons is obtained from the corresponding corrected ECAL and HCAL energy.

For the forward detectors, HF, CASTOR, and ZDC, there is no tracking information,
 therefore information from the calorimeter towers
only
is used for the analysis. The two HF calorimeters are each segmented into 13 rings in $\eta$. For this
analysis, the first two rings, covering $3.00 < \abs{\eta}< 3.15$, are excluded since they are partially located in the shadow of the
endcap calorimeter.
The subsequent ten rings of width
$\delta \eta=0.175$
are
grouped into 5 pairs of consecutive rings.
The last ring has a width of $\delta \eta=0.3$.
In total,  the transverse energy is measured in these  six $\eta$ bins in each HF calorimeter.
The calibration of the HF calorimeter is
derived from test beam data, and  radioactive sources
and has an accuracy of 10\%~\cite{Bayatian:2006jz}.
 The energy flow in the HF calorimeter is measured by summing  all energy deposits above
  the
  threshold
  of
 4\GeV
in a given ring.
Since CASTOR has no
$\eta$ segmentation, all energy deposits within it are summed together.  The absolute calibration of
the CASTOR calorimeter is achieved by a combination of extrapolation from the HF region
for 7\TeV  $\Pp\Pp$ data and simulation-based corrections. The accuracy of the energy scale is estimated to be 22\%.  The calibration of the ZDCs is based on electromagnetic interactions that produce single neutrons in the calorimeters with the energy $E_\text{beam}/A$~\cite{Grachov:2006ke}.

\section{Data taking and event selection}
\label{Sec:EventSelection}

The data for this analysis were recorded during the CERN LHC 2013 \pPb and \Pbp  data taking. During these runs,
31\nbinv of data were collected by CMS, of which
1.14\nbinv
 are used for this analysis. For this luminosity the statistical uncertainties  on the data are very small compared to the systematic ones.
For this paper
 the proton-going direction is defined to be towards positive rapidity, which implies that negative $\eta$ is in the lead-going direction.
 The switch in the  proton and lead beam directions
allows the use of CASTOR for measuring  \ET on both the lead- and
proton-going sides of the collision.
For this analysis, events are selected with an unbiased hardware
trigger requiring only the presence of proton and lead bunches in the CMS detector.
These bunches are detected by induction counters placed 175\unit{m} from the interaction point on each side of the experiment.
Furthermore, the presence of at least one single
reconstructed charged-particle track with $\abs{\eta}<2.4$ and $\pt>400\MeVc$ is required.
An offline selection reduces events from beam-gas or electromagnetic interactions~\cite{Khachatryan:2016xdg}.  Events are required to have
at least one HF calorimeter tower with more than 3\GeV of total energy on both the
positive and negative sides of the interaction point and at least
one reconstructed primary vertex with at least two associated tracks.
The effect of noise on the \ET measurement is estimated from a sample of events collected with a random trigger when no beams are present.

\section{Event centrality}
\label{Sec:Centrality}

In heavy ion collisions the activity or violence of a collision can be classified by several theoretical constructs~\cite{Busza:2018rrf}: the number of
nucleons that participate in the collision,  $N_\text{part}$, by the number of collisions between
participants, $N_\text{coll}$, and by the closest distance between the centers of the colliding nuclei, which is called the impact parameter, $b$.   The term centrality is used as an estimator of the impact parameter of the collisions. It is generally defined in terms of the multiplicity of charged-particles  or the
\ET produced in a given $\eta$ region.
 While in Monte Carlo (MC) simulations   $N_\text{part}$,  $N_\text{coll}$, and b
 are known, in
data, these variables cannot
be measured directly.
These quantities are estimated using  $\ET$ or
charged-particle multiplicity, which are both believed to scale monotonically with $N_\text{part}$ or $b$.

The centrality of a particular event  is defined to be the
percentile of events with values of the estimator larger than for that
particular event. A Glauber model is then  used to relate the centrality
to $N_\text{part}$,  $N_\text{coll}$, and $b$~\cite{Miller:2007ri}.

For symmetric heavy ion collisions the correlation of centrality with $N_\text{part}$
is strong~\cite{Chatrchyan:2012mb}, but for the much smaller \pPb system the fluctuations of $N_\text{part}$ with a
given experimental observable are large~\cite{Chatrchyan:2014hqa}.  For this paper
three different measures of centrality are investigated:
\begin{itemize}
  \item HF-Single: $\ET$ deposited in the Pb-going side of HF,
 in  $-5.0<\eta <-4.0$,
  \item HF-Double: The sum of $\ET$ deposited in both sides of HF,  in $4.0<\abs{\eta} < 5.0$,
  \item  $N_\text{track}$: number of  reconstructed
  tracks with
  $\pt>400\MeVc$
  and $\abs{\eta} < 2.4$.
\end{itemize}
When using the  charged-particle multiplicity or $\ET$ in given $\eta$ regions to define centrality there is an obvious autocorrelation between the centrality and
the  multiplicity or $\ET$ in that region.  It is not known, however, how far these correlations
extend over
 larger $\eta$ regions.
 The
near hermetic coverage of the CMS calorimeters, 13.2 units of $\eta$,  allow  for the most  complete picture of energy production  yet performed for proton-lead collisions at the LHC.
In order to understand the correlation that can arise from a choice of the centrality variable,
a study needs to be made over a large pseudorapidity range
for several centrality classes.

\section{Data analysis}

The measured transverse energy densities are
are presented for $\abs{\eta}<2.0$
in the tracker
 region, for $3.15<\abs{\eta}<5.20$ in the HF calorimeter, and for $5.2<\abs{\eta}<6.6$ in the
CASTOR calorimeter.  Because of  a switch of the beam direction during the data
taking, the CASTOR calorimeter can be used for both positive and negative $\eta$.

The transverse energy density  is calculated using  the following equation
\begin{linenomath*}
    \begin{equation}
       \frac{\rd\ET}{\rd\eta}(\eta)=\frac{C(\eta)}{N \Delta \eta}
    \sum_{j} \ET^{j} ( \text{if}~\ET^{j} > \text{noise}),
    \label{Eqn:EtCal}
    \end{equation}
\end{linenomath*}
where $N$ is the number of good events
that pass the online and the offline event selection,
 $C(\eta)$ is a correction factor that accounts for the reconstruction and triggering inefficiencies, and the index $j$ in the summation runs over all reconstructed particle-flow objects.
The correction is deduced from simulations and  is defined as
\begin{linenomath*}
   \begin{equation}
   C(\eta)=\frac{\sum_k \ET^k ( \text{generated})}{\sum_j \ET^j ( \text{reconstructed})( \text{if} ~\ET^{j} > \text{noise})},
   \label{Eqn:CorrectionFactor}
   \end{equation}
\end{linenomath*}
where the index $k$ in the top summation runs over all generated particles.
Using this definition $C(\eta)$ corrects the data
 from the detector level of the data to the stable-particle level,  \ie,  those particles with lifetimes $c\tau>1\unit{cm}$. This correction accounts for  the nonlinearity of the calorimeter response and the noise thresholds.
 The correction factor depends on the particle mix and average transverse momentum of the particles.
The \textsc{epos-lhc}, \textsc{hijing} and,  \textsc{qgsjet ii}  generators  are used to estimate $C(\eta)$.
For the analysis of the reconstructed simulated events, the event selection and
noise reduction requirements
are the
same as for the data analysis.  Events are selected by requiring at least one stable
particle
to be within the  HF $\eta$ range, $3.2<\abs{\eta}<5.2$, on  both sides.

In order to focus on the centrality dependence of the transverse energy as a function of $\eta$,
the events
are divided  into 10 bins of centrality, 0--10\%, 10--20\%, \etc.
Here we
consider 0--10\% to be \emph{central} and any other centrality to be
\emph{peripheral}.
Using these definitions the ratio of peripheral to central
$\rd\ET/\rd\eta$
is defined as
\begin{linenomath*}
  \begin{equation}
S_\text{PC} (\eta)=\frac{\frac{\rd\ET}{\rd\eta}(\text{peripheral},\eta)}{\frac{\rd\ET}{\rd\eta}(\text{central},\eta)}.
\label{Eqn:SpcDef}
\end{equation}
This can be written as
\begin{equation}
S_\text{PC}(\eta)=\frac{\sum_{i} \ET^{i} (\text{peripheral})}
{\sum_{i} \ET^{i} (\text{central})}\frac{N_\text{peripheral}} {N_\text{central}}\frac{C(\text{peripheral},\eta)} {C(\text{central},\eta)}.
\label{Eqn:SpcReduced}
  \end{equation}
\end{linenomath*}
Since $S_\text{PC}$  represents a ratio of results for two data samples multiplied by a ratio of two correction factors, correlated uncertainties tend to cancel, which is a major advantage of this approach.
This method of studying the centrality dependence, rather than the more traditional ratio of central to peripheral events, exploits the fact that the 0--10\% centrality class has the smallest fractional uncertainties and so minimizes the correlated uncertainties when comparing data from different centrality classes.

\section{Systematic uncertainties}

In this analysis, there are several sources of systematic uncertainties on $\rd\ET/\rd\eta$:
\begin{enumerate}
\item The differences in $\ET$ spectra and particle composition between data and the MC simulation used to generate correction factors. The impact of these differences is  estimated by generating MC samples with different particle mixes and \ET spectra. These effects are most important in the tracker, $\abs{\eta} < 2.4$,  and HF  regions, $3.15<\abs{\eta} < 5.20$,  and are less than 3\%.
\item Uncertainties in the calorimeter energy scale. These are estimated by the differences in calibration from various methods. These contribute less than 1\% in the tracker region, 10\% for HF, and 22\% for CASTOR.
\item Method of handling the noise in the calorimeters. These uncertainties are estimated by using different sets of noise reduction requirements  in the analysis.  These uncertainties are less than 3\% in the tracker and HF regions, and  are negligible for CASTOR.
\item Any asymmetries between the positive and negative sides of CMS, e.g., from dead channels, \etc. The data from \pPb collisions at a given positive $\eta$ are  compared to those of \Pbp events at the corresponding negative $\eta$. These uncertainties  are up to 5.0\% in the tracker region,  and up to 3.5\% in the HF region.
\end{enumerate}
The uncertainties described above  are evaluated separately
in the tracker, HF, and CASTOR regions and summed in quadrature.  For the CASTOR region the uncertainty in  the energy scale dominates the total systematic uncertainty.
Table \ref{Tab:SysEtTracker}
lists  the  systematic uncertainties on $\rd\ET/\rd\eta$ and $S_\text{PC}$  for each $\eta$ region as a function centrality as defined by HF-Double.
The systematic uncertainties are the smallest for the most central events.
 For  $S_\text{PC}$, there is a high degree of cancellation between the uncertainties in different centrality classes. In particular the energy scale  and  forward/backward systematic uncertainties cancel almost completely
  while the uncertainties related to the
  simulation and noise reduction only partially cancel.
 The net result is that the  systematic uncertainties in $S_\text{PC}$ are considerably smaller than those in \ET.

\begin{table}[h]
\centering
\topcaption{Systematic uncertainties  in $\rd\ET/\rd\eta$ and $S_\text{PC}$
for the tracker region,
the HF region,
and the CASTOR region as a function of centrality defined by HF-Double.
The $S_\text{PC}$ ratio is by construction unity for 0 - 10\% centrality and is not defined for minimum bias events. }
\begin{scotch}{ccccccc}
 \multicolumn{1}{c}{ }  & \multicolumn{3}{c}{$\rd\ET/\rd\eta$ systematic (\%)}  & \multicolumn{3}{c}{$S_\text{PC}$ systematic (\%)}\\
Centrality        &  Tracker & HF & CASTOR  &  Tracker & HF & CASTOR \\ \hline
0--10\%        &  3.7 &  10.1 & 22   & \multicolumn{3}{c}{\NA} \\
10--20\%      &  3.8 &  10.1 & 22  & 1.0 & 1.1  & 1.3 \\
20--30\%      &  3.8 &  10.1 & 22  & 1.3 & 1.1  & 1.5 \\
30--40\%      &  3.8 &  10.1 & 22  & 1.3 & 1.2  & 4.1 \\
40--50\%      &  4.2 &  10.1 & 22  & 1.3 & 1.2  & 4.1 \\
50--60\%      &  4.5 &  10.1 & 22  & 1.3 & 1.2  & 4.1 \\
60--70\%      &  5.1 &  10.2 & 22  & 1.6 & 1.3  & 4.1\\
70--80\%      &  7.0 &  10.4 & 23  & 3.5 & 1.3  & 4.1\\
Min. bias        &  4.2 &  10.1 & 22  & \multicolumn{3}{c}{\NA} \\
\end{scotch}
\label{Tab:SysEtTracker}
\end{table}

\section{Results}
\begin{figure}[hbtp]
\centering
\includegraphics[width=1.0\columnwidth]{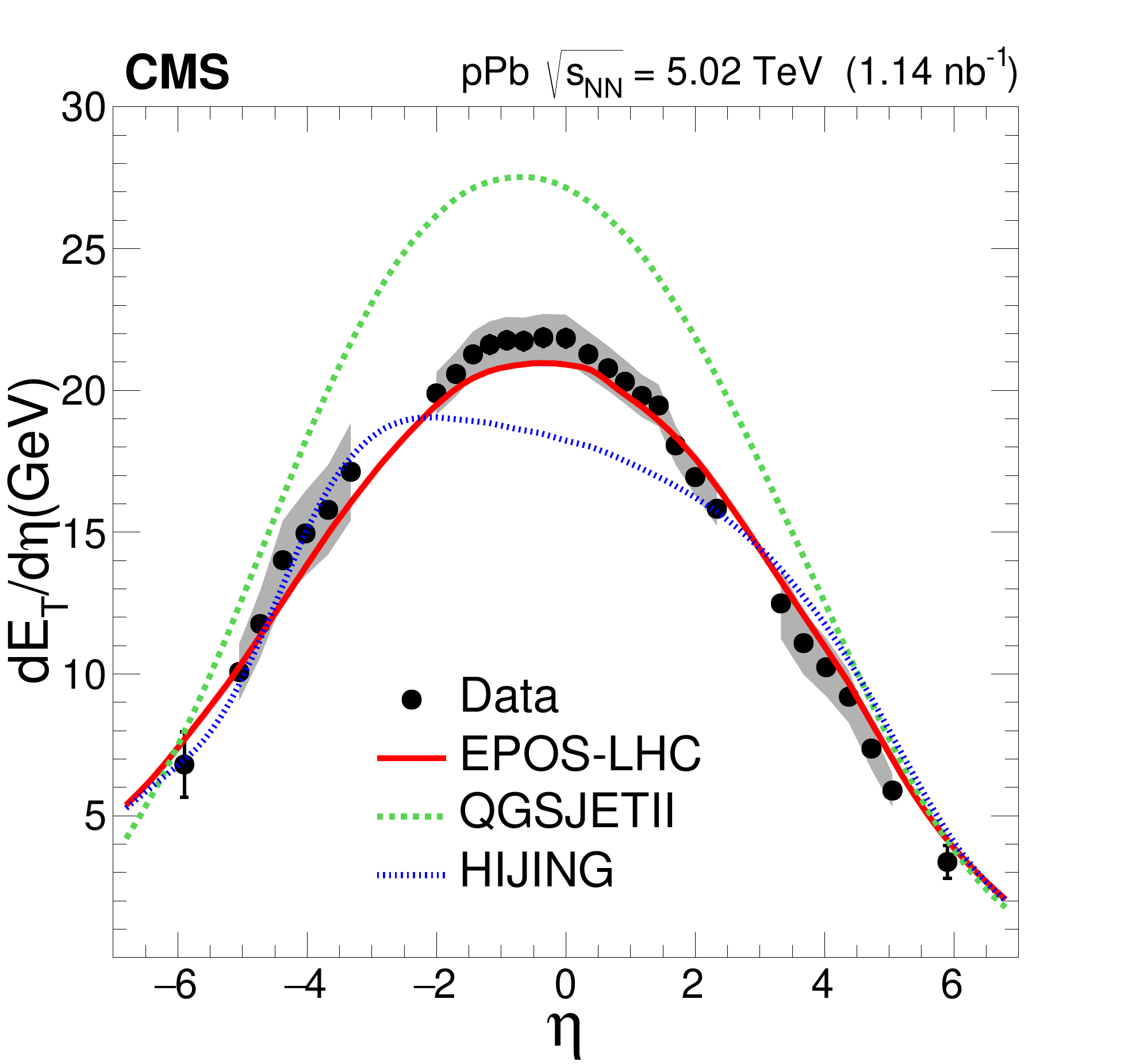}
\caption{Transverse energy density
versus $\eta$ from minimum bias \pPb
collisions at  .
at $\sqrtsNN=5.02\TeV$.
The proton is moving towards positive $\eta$.  The statistical uncertainties are smaller than
the size of the data points and the total errors are dominated by the systematics. The systematic uncertainties are largely
correlated point to point within the central  and with the HF regions  and so shown by gray bands there. The systematic uncertainties for the most forward and backward data points \ie $\eta=\pm5.9$ are uncorrelated with those of central and HF regions and so are shown as vertical bars.   Predictions from the \textsc{epos-lhc} (red solid), \textsc{qgsjet ii} (green dashed), and \textsc{hijing} (blue dotted) event generators are also shown.
\label{Corrected_minbias_data}}
\end{figure}

The most basic measurement of \ET production is performed for
the minimum bias selection as a function of $\eta$.
Figure \ref{Corrected_minbias_data}  shows the
resulting
$\rd\ET/\rd\eta$ versus $\eta$  for data and
 for
 predictions from
 the \textsc{epos-lhc}, \textsc{qgsjet ii} and \textsc{hijing} models.
 The \textsc{hijing} event
generator is based on a two-component model for hadron production in high-energy nucleon and nuclear
collisions. Hard parton scattering is assumed to be described by perturbative QCD, and soft
interactions are approximated by string excitations with an effective cross section. For heavy nuclei,  initial parton distributions
are modified with respect to  those of free protons. Also,  multiple scatterings inside a nucleus
lead to transverse momentum
($\pt$)
broadening of both initial- and final-state partons.  Both the \textsc{epos-lhc} and \textsc{qgsjet ii} models use
Gribov--Regge
theory to give a self consistent quantum mechanical treatment of the initial  parton-level interactions  without an arbitrary division into soft and hard interactions~\cite{Drescher:2000ha}.
The \textsc{epos-lhc} generator also includes
 a phenomenological implementation of gluon
saturation.
After the initial interactions, this model uses a hydrodynamic approach to
evolve regions of high energy density.
The \textsc{qgsjet ii} generator allows parton cascades to split and
merge via pomeron-pomeron interactions,
but does
not include a hydrodynamic component.
Saturation effects are produced via higher-order pomeron-pomeron
interactions.

From Fig.~\ref{Corrected_minbias_data}  it can be seen
that
$\rd\ET/\rd\eta |_{\eta=0}\approx  22\GeV$.
This is 1/40
of the value observed for the $2.5\%$ most central PbPb collisions~\cite{Chatrchyan:2012mb}. However,
since  the cross sectional area of a \pPb collision  is much smaller than that of a
central PbPb collision~\cite{Krane,ANGELI201369}, this result  implies that the maximum energy density in \pPb collisions is comparable to that achieved in
PbPb collisions.

By comparing $\rd\ET/\rd\eta$ to $\rd N_\text{ch}/\rd\eta$,
which was previously measured by our experiment in proton-lead collisions at the
same energy~\cite{Sirunyan:2017vpr}, it is possible to calculate the transverse energy per charged-particle. At the center-of-mass pseudorapidity
  we find $\ET/N_\text{ch}=1.31\pm0.07\GeV/\text{particle}$  for minimum bias  \pPb collisions at
  $\sqrtsNN=5.02\TeV$.
  This is somewhat higher than the value of
  $1.0\pm0.1\GeV/\text{particle}$ reported by PHENIX for \dAu collisions at
  $\sqrtsNN=200\GeV$~\cite{Adler:2004zn}.

Predictions from the \textsc{epos-lhc} model  are close to
the data over the entire pseudorapidity range
while
those from the \textsc{hijing}  model
are  consistent with the data for $\eta<-3$ and $\eta>2$, but are significantly below the data at midrapidity,
\ie, $\abs{\eta}<2$.  Predictions from the  \textsc{qgsjet ii} generator are consistently above the data over the entire $\eta$ range.
The peak of the data distribution is around $\eta=-0.5$.
Both \textsc{epos-lhc} and \textsc{qgsjet ii} generators peak close to this value while \textsc{hijing} has a maximum at $\eta=-2.5.$

\begin{figure}[hbtp]
\centering
\includegraphics[width=1.0\columnwidth]{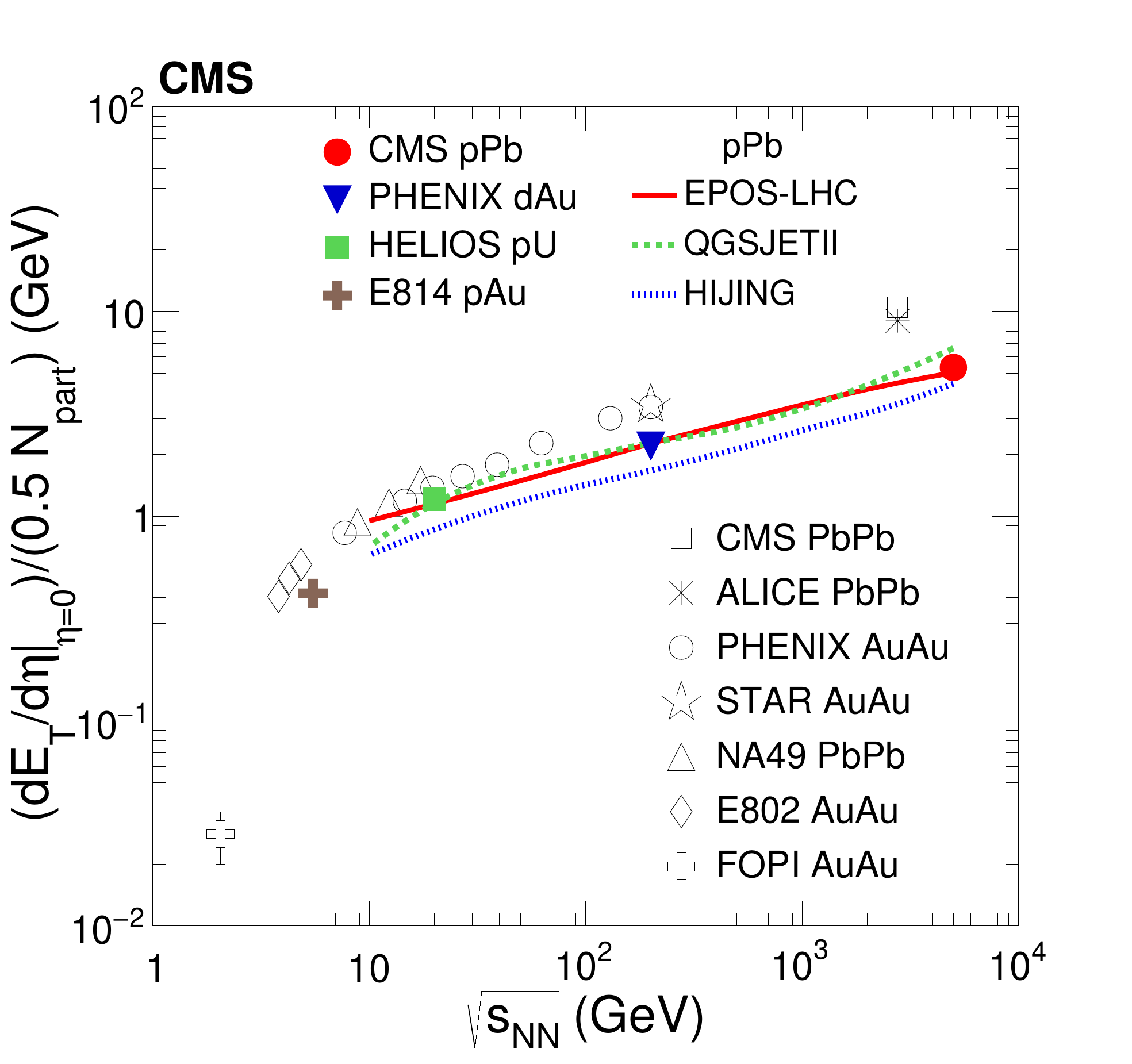}
\caption{Transverse energy density per participating nucleon-nucleon pair evaluated at
 $\eta_\text{cm}$
versus $\sqrtsNN$ for minimum bias  \pAu, \pU, \dAu, and \pPb collisions.
For the CMS \pPb data at
$\sqrtsNN = 5.02\TeV$, $N_\text{part}$ was estimated to be $8.0 \pm 0.2$  using the method described in \cite{Miller:2007ri}.
The uncertainties are generally smaller than the size of the data points.
Also shown are the corresponding results for  central AuAu and  PbPb collisions, as well as simulation for minimum bias \pPb collisions from three event
generators~\cite{Akesson:1992uv,Chatrchyan:2012mb,Adam:2016thv,Adler:2004zn,Abbott:2001gd,Bachler:1999hu,Afanasiev:2002fk,Afanasiev:2002mx,Reisdorf:1996qj,Pelte:1997rg,Hong:2001tm,Star:2004,E802:1999}.
\label{EtNpartRootS}}
\end{figure}

Figure \ref{EtNpartRootS} shows the  transverse energy density at midrapidity,
$\rd\ET/\rd\eta |_{\eta=0}$, versus  $\sqrtsNN$ for minimum bias \pA and \dA collisions for
several experiments~\cite{Akesson:1992uv, Abbott:2001gd,Adare:2015bua}.
The data are averaged over a small region around the center-of-mass pseudorapidity,
with a typical $\abs{\eta-\eta_\text{cm}}<0.5$.
  To account for the different system sizes the $\rd\ET/\rd\eta$
values are normalized to the number of participating pairs of nucleons in the collisions. For the CMS data
$N_\text{part}$ was estimated to be $8.0 \pm 0.2$  using the method described in \cite{Miller:2007ri}.
Figure~\ref{EtNpartRootS} also shows a compilation of results for central AA collisions from Ref.~\cite{Adare:2015bua} with the addition of a recent
 ALICE PbPb data point~\cite{Adam:2016thv}.
 Although the geometries and lifetimes of \pA and AA collisions are very different, it is interesting to note that the  \pPb minimum bias value of   $5.33\pm$0.25\GeV per participant pair    is
higher than the central AuAu result at $\sqrtsNN=200\GeV$~\cite{Adare:2015bua} and  consistent with the
peripheral PbPb result at
2.76\TeV~\cite{Chatrchyan:2012mb}.

The rate of increase of $\rd\ET/\rd\eta |_{\eta=0}$ with $\sqrtsNN$ is stronger for AA than for \pA collisions. This is  expected because of the
increased stopping power, \ie, the ability to decelerate nucleons, of   heavy nuclei compared to
protons~\cite{Videbaek:1995mf,Busza:1983rj}.
The stopping power controls the total amount of energy available for particle production.
 The rapidity shift of the incoming nucleons is proportional to the beam rapidity for energies up to
  $\sqrtsNN=63\GeV$, but then seems to
  saturate~\cite{Busza:1983rj,Bearden:2003hx,Arsene:2009aa,Wohrmann:2013nta}.  This limit to the deceleration may be the reason for the change in slope of the AA data near  $\sqrtsNN \approx 10\GeV$.
The \pA data also seems to change slope in this region but  unfortunately the sparsity of data with
 $\sqrtsNN$ between 5 and 20\GeV make it difficult to determine where this change happens in \pA collisions.

For energies
above  $\sqrtsNN \approx 10\GeV$ the scaled transverse energy density
increases as  a power law
according to
$s_\text{NN}^\gamma$.
Such an energy dependence has been previously observed for the charged-particle multiplicity density,
$\rd N^\pm/\rd\eta$,
 near $\eta=0$~\cite{Adare:2015bua,Adam:2016thv,Sirunyan:2017vpr}.
 Table \ref{Tab:PowerLaw} lists the results of fitting the energy dependence of
the scaled $\rd N^\pm/\rd\eta$
 and  $\rd\ET/\rd\eta$
 for central events to a function of  the form $s_{\text{NN}}^\gamma$.  The \ET rises more rapidly with energy than the charged-particle multiplicity. Again this is expected because  the mean transverse momentum is also increasing
 with beam energy~\cite{Adamczyk:2017iwn}.  This difference in the energy dependence of \ET and multiplicity production is stronger for AA than for \pA collisions.
  This suggests that the mean transverse momentum rises faster with energy in  AA than in \pA collisions.
 \begin{table}[h]
 \centering
 \topcaption{Values of exponents from fitting the energy dependence of
 $\rd N^\pm/\rd\eta$~\cite{Sirunyan:2017vpr} and
 $\rd\ET/\rd\eta$ at midrapidity
 to a function of the form
 $s_{\text{NN}}^\gamma$  for minimum bias proton-nucleus and central nucleus-nucleus collisions.}
\begin{scotch}{ccc}
Collision   &  $\gamma$ for N$_\text{ch}$ & $\gamma$ for \ET \\  \hline
pA                & $0.103 \pm .005$ &   $0.135   \pm .003$ \\
AA                & $0.158 \pm .004$ &   $0.205   \pm .005$ \\
\end{scotch}
\label{Tab:PowerLaw}
\end{table}

Figure \ref{EtNpartRootS} also shows simulations of \pPb interactions  at various energies.
Predictions from the  \textsc{epos-lhc} model are consistent with  the  data from
$\sqrtsNN=20\GeV$ to 5.02\TeV.
The  \textsc{qgsjet}  model is consistent with the  20 and 200\GeV data, but is somewhat higher than the data at
$\sqrtsNN=5.02\TeV$.
The \textsc{hijing} generator has a similar energy dependence of the data, but is consistently below the experimental results.

\begin{figure*}[hbtp]
\centering
\includegraphics[width=1.0\textwidth]{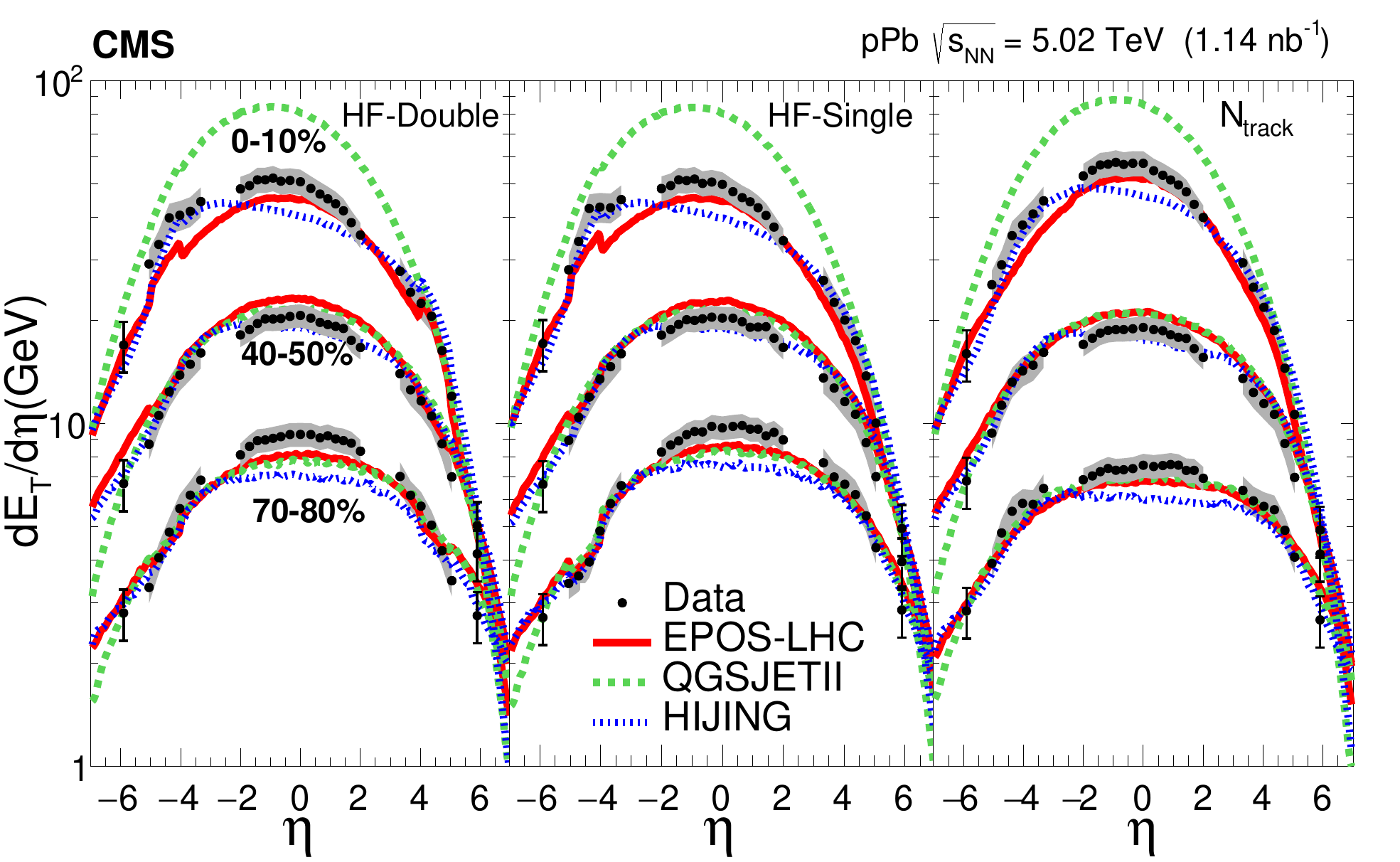}
\caption{Transverse energy density
 versus $\eta$ and centrality  from 5.02\TeV \pPb collisions for the
HF-Double (left), HF-Single (center), and $N_\text{track}$ (right) centrality definitions for data and for predictions from the \textsc{epos-lhc}, \textsc{qgsjet ii},  and \textsc{hijing} event generators,  for  0--10\%  (upper), 40--50\% (middle), and 70--80\% (lower) central collisions. The uncertainties are dominated by the systematic components, which are largely correlated point-to-point in the central region and in HF, and which are shown by gray bands there.
  \label{Eflow_Centrality_Comparison}}
\end{figure*}

Figure \ref{Eflow_Centrality_Comparison}  shows
$\rd\ET/\rd\eta$
versus $\eta$ for \pPb collisions at $\sqrtsNN=5.02\TeV$
for several centralities and for three different definitions of centrality
for both data and simulations.
For 0--10\% most central collisions,
$\rd\ET/\rd\eta |_{\eta=0}$  exceeds 50\GeV.
For the top 10\% central \pPb collisions it is reasonable to assume a complete overlap of the incoming proton with the lead nucleus.  Thus, the transverse area $A_\perp$ corresponds to   the total proton-proton (\Pp\Pp)  cross section, $\sigma_{\Pp\Pp}^\text{tot}$, at $\sqrt{s}=5.02\TeV$.
The TOTEM collaboration has measured
$\sigma_{\Pp\Pp}^\text{tot}$
at 2.76, 7, 8, and 13 TeV  \cite{Antchev:2013paa,Antchev:2016vpy,Antchev:2013iaa,Antchev:2017dia}.
Based on these results we estimate $\sigma_{\Pp\Pp}^\text{tot} = 94\pm1$ mb at $\sqrt{s} = 5.02 \TeV$.
Furthermore, the factor  $\rd y/\rd\eta$ needed for
Eq.~(\ref{Eqn:EnergyDensity})  depends upon the particle mix and \PT spectra. This factor is evaluated using simulated events from the three MC generators and  is found to be $1.12\pm0.03$.
With these considerations Eq.~(\ref{Eqn:EnergyDensity}) implies an energy density at a time
$\tau_0=1\unit{fm}/c$
of the order  of
$4.5\GeVfmcube$  for the top 10\% \pPb  collisions.
This is above the expected threshold for  the production of a quark-gluon plasma estimated from lattice
QCD
calculations~\cite{Karsch:2001cy}.

For peripheral events the
peak of  $\rd\ET/\rd\eta$  is close to the nucleon-nucleon center-of-mass  pseudorapidity,  $\eta_\text{cm}=0.465$.
The peak moves  towards the Pb side as the centrality increases, reflecting the increased momentum from the lead-going nucleons.
For the most central events, the peak of  $\rd\ET/\rd\eta$ is at $\eta \approx -1.0$, \ie, 1.4  units
below
$\eta_\text{cm}$. This is very close to the pseudorapidity shift  observed for central \pU   collisions  at
 $\sqrtsNN=20\GeV$~\cite{Akesson:1992uv}, suggesting that the stopping power of heavy nuclei for protons is almost independent of the  center-of-mass energy for energies above 20\GeV.  For AA collisions a similar energy independence of the stopping power has been observed for $\sqrtsNN$ greater
 than 63\GeV~\cite{Bearden:2003hx,Arsene:2009aa,Wohrmann:2013nta}.

All three event generators  show a large increase of
$\rd\ET/\rd\eta |_{\eta=0}$ and a shift of $\langle \eta \rangle$ towards
the lead-going side as the
centrality increases.
However, for the 0--10\%, centrality selection the \textsc{hijing} distribution peaks at significantly lower $\eta$ than the data.
Predictions from the \textsc{epos-lhc} model are closest to the data for $\abs{\eta}<2$, whereas  the \textsc{hijing} generator gives a better description of the data in the lead-going region, \ie, $\eta<-3$. In the proton-going region, \ie, $\eta>3$,  the two generators are closer to each
other and the data.
The \textsc{qgsjet ii} predictions significantly exceed the data at all rapidities for the 0--10\% most
central collisions, but are close to the data for the 40--50\% and 70--80\% centrality classes.
As the centrality increases,  $\rd\ET/\rd\eta |_{\eta=0}$    increases faster for the
$N_\text{track}$
centrality definition than for the HF-Single or HF-Double definitions.
This effect results from the autocorrelation with the centrality definition.

\begin{figure}[hbtp]
\centering
\includegraphics[width=1.0\columnwidth]{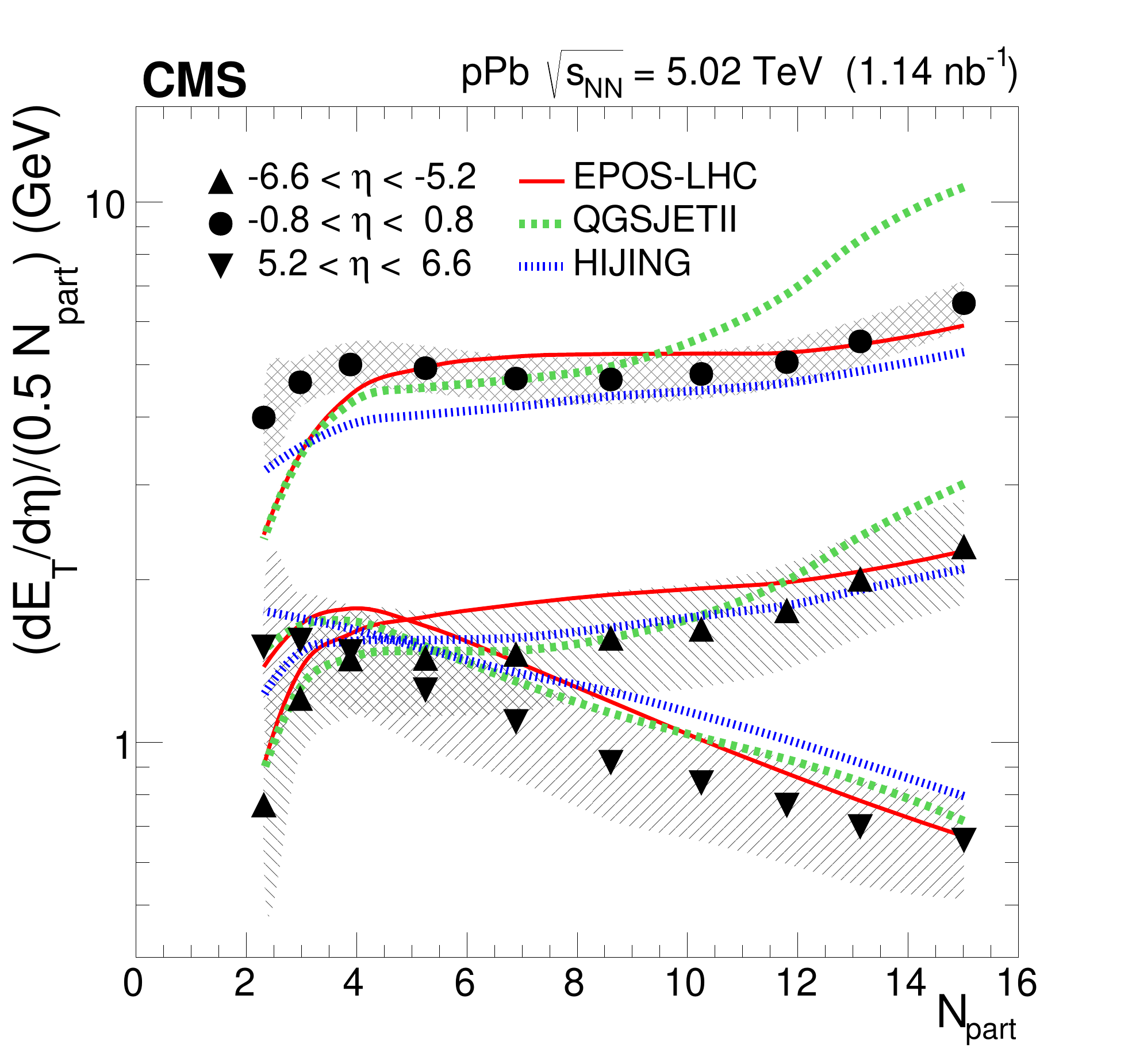}
\caption{Transverse energy density per participating nucleon-nucleon pair
versus $N_\text{part}$ for different $\eta$ ranges.
The $N_\text{part}$ values are based on  the method described in \cite{Miller:2007ri}.
The HF-Single method was used to define centrality.
The total experimental uncertainties are shown by gray bands.
The values of $N_\text{part}$
were calculated using the method described in \cite{Miller:2007ri}.
\label{detdeta_tonpart}}
\end{figure}

Figure \ref{detdeta_tonpart} shows  $\rd\ET/\rd\eta$ scaled by the number of participant nucleon pairs as a function
of $N_\text{part}$  for
the far lead-going region $-6.6<\eta<-5.2$,
the midrapidity  region $\abs{\eta}<0.8$,
and the far proton-going region $5.2<\eta<6.6$.  The centrality definition is  based on the HF-Single selection, \ie, $-5.0 < \eta < -4.0$.
It is clear that the centrality dependence of \ET production
varies strongly with $\eta$.
For $N_\text{part} > 3$ we find that
 $\rd\ET/\rd\eta$ per participant nucleon pair rises with $N_\text{part}$ in the lead-going and midrapidity  regions, but falls for the far proton-going region. This is consistent with the backward shift of the  mean $\eta$ with centrality observed in Fig.~\ref{Eflow_Centrality_Comparison}.

 Figure \ref{detdeta_tonpart} also shows  model predictions from \textsc{epos-lhc}, \textsc{qgsjet ii}, and \textsc{hijing}.
At midrapidity none of the generators is consistent with the data over the whole range of
$N_\text{part}$.   In particular, the \textsc{qgsjet ii} model has a much stronger centrality dependence than the data.
  For the lead-going   region
all three generators are consistent with the data within errors.
For the proton-going region,
all three generators are above the data, but  predictions from the \textsc{qgsjet ii} model are closer to the data than those from either \textsc{epos-lhc} or \textsc{hijing}.

\begin{figure*}[hbtp]
\centering
\includegraphics[width=1.0\textwidth]{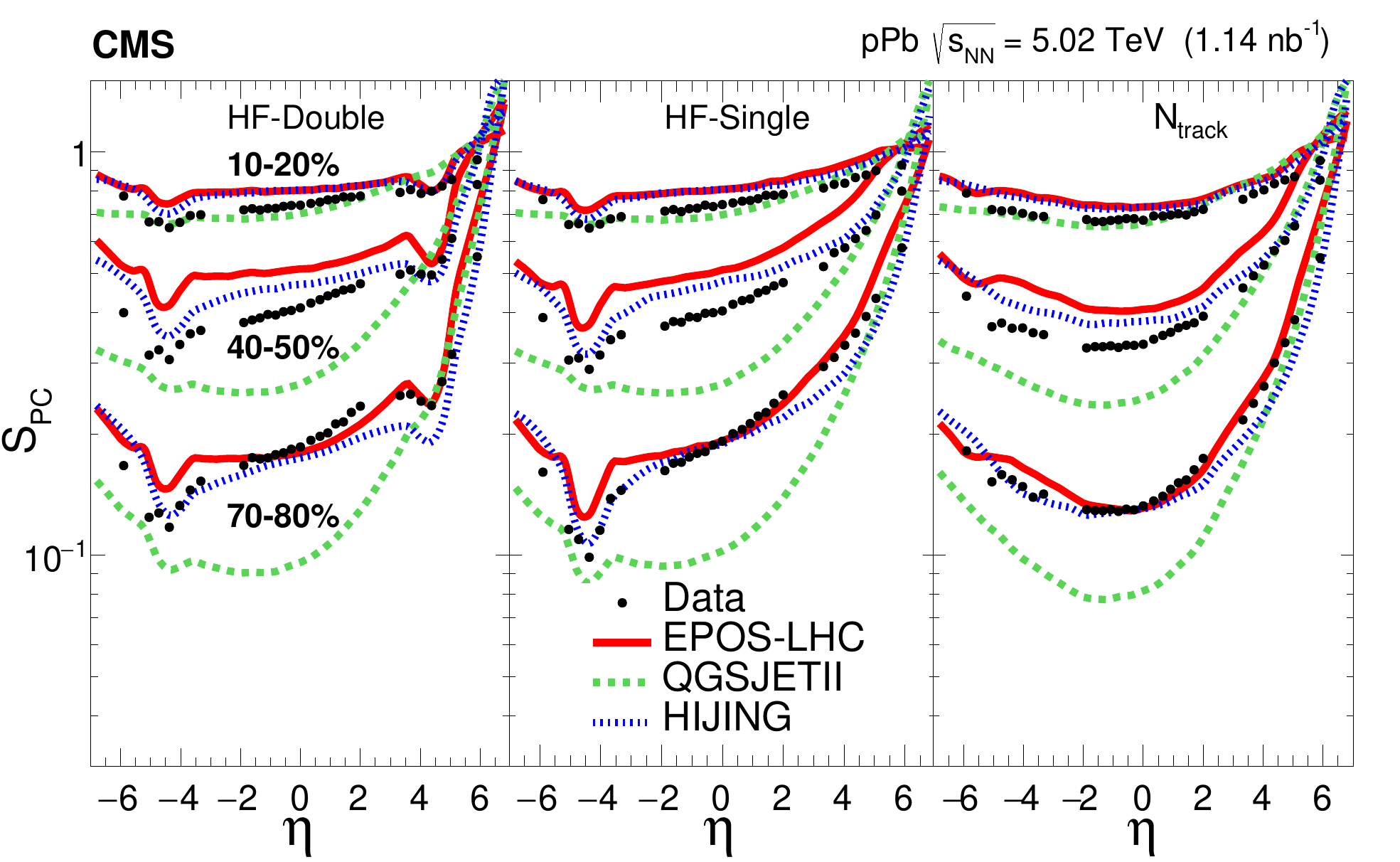}
\caption{Ratio of peripheral to central \ET production, $S_\text{PC}$,
 as a function of $\eta$ for three centrality ranges for HF-Double (left),
HF-Single (middle), and $N_\text{track}$ (right) for  data, and for the \textsc{epos-lhc}, \textsc{qgsjet ii}, and \textsc{hijing} event generators. The
systematic uncertainties  are dominant and are of comparable  size to the data points.
\label{SPC_centrality_log}}
\end{figure*}

Figure~\ref{SPC_centrality_log} shows
$S_\text{PC}$  as a function of $\eta$ for three centrality ranges and for all three centrality
definitions for data as well as for predictions from the  \textsc{epos-lhc}, \textsc{qgsjet ii}, and \textsc{hijing} event generators.
Note that as per the definition, for each centrality bin, say 40--50\%,   $S_\text{PC}$ shows the ratio of the
$\rd\ET/\rd\eta$ in that ``peripheral" bin to $\rd\ET/\rd\eta$ for the 0--10\% most central events.
As expected,
 $S_\text{PC}$  increases with centrality for all
centrality definitions.
The $S_\text{PC}$ value tends to rise with $\eta$ since the centrality dependence of
$\ET$ production is stronger on the lead-going side than on the proton-going side. This is presumably because particles moving in the lead direction are more likely to have multiple interactions than particles moving in the proton-going region.

The autocorrelation between the centrality definition and the measure of   $\rd\ET/\rd\eta$
suppresses $\rd\ET/\rd\eta$  for peripheral events and enhances it for central events in the $\eta$ region that is used for the centrality determination.
These two effects naturally induce a dip in the ratio of peripheral to central distributions in that particular $\eta$ region.
This effect is strongest for
$S_\text{PC}$ in the 70--80\% centrality class for the HF-Single and HF-Double centrality definitions.
While
the HF centrality is based on
$4<\abs{\eta} <5$, the impact of the autocorrelations is very clearly visible  over one to two more units of $\eta$.
 In contrast, the $N_\text{track}$ centrality definition uses all tracks with $\abs{\eta} < 2.4$,  resulting in
 a much smoother $S_\text{PC}$ as a  function of $\eta$.

The \textsc{qgsjet ii} model gives the best description of $S_\text{PC}$ in the 10--20\% centrality range, however, it significantly
underestimates the magnitude of $S_\text{PC}$ in all other cases,
implying that it significantly overestimates the increase of $\rd\ET/\rd\eta$ with centrality.  The \textsc{hijing} and \textsc{epos-lhc} generators in general do a better job in describing the magnitude of $S_\text{PC}$
with \textsc{epos-lhc}, giving the best description in the 70--80\% centrality range.
 None of the models gives a complete description of the centrality dependence of the data.

The \textsc{qgsjet ii} generator also underestimates the  dips in $S_\text{PC}$ as a function of $\eta$
 for both the HF-Double and HF-Single definitions, of centrality.
This is most clearly seen  for the HF-Double definition in the forward region where the data show significant dips but the
\textsc{qgsjet ii} distributions  increase monotonically with $\eta$.
The \textsc{hijing} and \textsc{epos-lhc} models both produce dips in the same $\eta$ regions as the data for both HF centrality
definitions  but neither generator is able to predict the shape of $S_\text{PC}$ over the full $\eta$ range.
This failure to reproduce the  $\eta$ dependence  of  $S_\text{PC}$ suggests that the generators do not correctly model the correlations present in proton-lead collisions.

\section{Summary}

In this paper we  report  the centrality and pseudorapidity ($\eta$) dependence of  transverse energy (\ET) production from
\pPb  collisions at $\sqrtsNN=5.02\TeV$
 over 13.2 units of $\eta$.
The \ET per participant pair in minimum
bias \pPb events at $\sqrtsNN=5.02\TeV$ is comparable to that of peripheral  PbPb collisions
at  2.76\TeV.
 At midrapidity the energy density at a proper time $\tau_0=1\unit{fm}/c$
  is of order of
  4.5\GeVfmcube
for the top 10\%  most central \pPb  collisions, which
is comparable to those observed in PbPb collisions.
 As the centrality of the collision increases, the total
\ET increases dramatically and the  mean $\eta$ of the \ET distribution moves towards the
lead-going side of the collision.   For central collisions, the peak of
$\rd\ET/\rd\eta$ is 1.4 units below the center-of-mass pseudorapidity. This pseudorapidity shift
is almost the same as for \pU   collisions at $\sqrtsNN=20\GeV$.

The \textsc{epos-lhc} event generator gives a good description of the minimum bias
$\rd\ET/\rd\eta$ distribution and  peaks at an $\eta$ value close to that of the data for all
centralities.
The centrality dependence of  \ET production for \textsc{qgsjet ii} is  stronger than that of the data. This model is below the data for
70--80\% peripheral events and almost a factor of two above the data for the 10\% most central events.
Near midrapidity the  \textsc{hijing} generator tends to
underestimate
the magnitude of  $\rd\ET/\rd\eta$ and  for central collisions predicts a peak that is at significantly lower $\eta$ than in
the data.

Similarly to what has been seen in particle production at lower energy~\cite{Busza:2004mc},
the $\rd\ET/\rd\eta$
 per participating nucleon-nucleon pair  increases with  the number of nucleons that participate in the collisions ($N_\text{part}$)  for $\eta$ values on the lead side;
it is rather independent of $N_\text{part}$ near midrapidity;
and it decreases  with $N_\text{part}$ for $\eta$ values  on the proton side.
The $\eta$ region used to define centrality has a strong
impact
on the nature of the events selected.
There is a significant autocorrelation of
the $\eta$ range used to define centrality with
$\rd\ET/\rd\eta$
 both for data,
and the \textsc{epos-lhc},
\textsc{qgsjet ii} and \textsc{hijing} event generators.   None of the tested event generators are able to capture all
aspects of the autocorrelations seen in data.

It is clear that  cosmic ray event generators have difficulties modeling both the centrality and $\eta$ dependence of  proton-lead collisions.
While the proton-lead system
 is significantly larger than the  proton-nitrogen and proton-oxygen
collisions occurring in air showers,
 these data
illustrate the need for a better understanding of nuclear effects. Ultimately, protons
colliding with light nuclei would be  most valuable for this purpose.

\begin{acknowledgments}
We congratulate our colleagues in the CERN accelerator departments for the excellent performance of the LHC and thank the technical and administrative staffs at CERN and at other CMS institutes for their contributions to the success of the CMS effort. In addition, we gratefully acknowledge the computing centers and personnel of the Worldwide LHC Computing Grid for delivering so effectively the computing infrastructure essential to our analyses. Finally, we acknowledge the enduring support for the construction and operation of the LHC and the CMS detector provided by the following funding agencies: BMBWF and FWF (Austria); FNRS and FWO (Belgium); CNPq, CAPES, FAPERJ, FAPERGS, and FAPESP (Brazil); MES (Bulgaria); CERN; CAS, MoST, and NSFC (China); COLCIENCIAS (Colombia); MSES and CSF (Croatia); RPF (Cyprus); SENESCYT (Ecuador); MoER, ERC IUT, and ERDF (Estonia); Academy of Finland, MEC, and HIP (Finland); CEA and CNRS/IN2P3 (France); BMBF, DFG, and HGF (Germany); GSRT (Greece); NKFIA (Hungary); DAE and DST (India); IPM (Iran); SFI (Ireland); INFN (Italy); MSIP and NRF (Republic of Korea); MES (Latvia); LAS (Lithuania); MOE and UM (Malaysia); BUAP, CINVESTAV, CONACYT, LNS, SEP, and UASLP-FAI (Mexico); MOS (Montenegro); MBIE (New Zealand); PAEC (Pakistan); MSHE and NSC (Poland); FCT (Portugal); JINR (Dubna); MON, RosAtom, RAS, RFBR, and NRC KI (Russia); MESTD (Serbia); SEIDI, CPAN, PCTI, and FEDER (Spain); MOSTR (Sri Lanka); Swiss Funding Agencies (Switzerland); MST (Taipei); ThEPCenter, IPST, STAR, and NSTDA (Thailand); TUBITAK and TAEK (Turkey); NASU and SFFR (Ukraine); STFC (United Kingdom); DOE and NSF (USA).

\hyphenation{Rachada-pisek} Individuals have received support from the Marie-Curie program and the European Research Council and Horizon 2020 Grant, contract No. 675440 (European Union); the Leventis Foundation; the A. P. Sloan Foundation; the Alexander von Humboldt Foundation; the Belgian Federal Science Policy Office; the Fonds pour la Formation \`a la Recherche dans l'Industrie et dans l'Agriculture (FRIA-Belgium); the Agentschap voor Innovatie door Wetenschap en Technologie (IWT-Belgium); the F.R.S.-FNRS and FWO (Belgium) under the ``Excellence of Science - EOS" - be.h project n. 30820817; the Ministry of Education, Youth and Sports (MEYS) of the Czech Republic; the Lend\"ulet (``Momentum") program and the J\'anos Bolyai Research Scholarship of the Hungarian Academy of Sciences, the New National Excellence Program \'UNKP, the NKFIA research grants 123842, 123959, 124845, 124850 and 125105 (Hungary); the Council of Science and Industrial Research, India; the HOMING PLUS program of the Foundation for Polish Science, cofinanced from European Union, Regional Development Fund, the Mobility Plus program of the Ministry of Science and Higher Education, the National Science Center (Poland), contracts Harmonia 2014/14/M/ST2/00428, Opus 2014/13/B/ST2/02543, 2014/15/B/ST2/03998, and 2015/19/B/ST2/02861, Sonata-bis 2012/07/E/ST2/01406; the National Priorities Research Program by Qatar National Research Fund; the Programa Estatal de Fomento de la Investigaci{\'o}n Cient{\'i}fica y T{\'e}cnica de Excelencia Mar\'{\i}a de Maeztu, grant MDM-2015-0509 and the Programa Severo Ochoa del Principado de Asturias; the Thalis and Aristeia programs cofinanced by EU-ESF and the Greek NSRF; the Rachadapisek Sompot Fund for Postdoctoral Fellowship, Chulalongkorn University and the Chulalongkorn Academic into Its 2nd Century Project Advancement Project (Thailand); the Welch Foundation, contract C-1845; and the Weston Havens Foundation (USA).

\end{acknowledgments}

\bibliography{auto_generated}

\cleardoublepage \appendix\section{The CMS Collaboration \label{app:collab}}\begin{sloppypar}\hyphenpenalty=5000\widowpenalty=500\clubpenalty=5000\vskip\cmsinstskip
\textbf{Yerevan Physics Institute, Yerevan, Armenia}\\*[0pt]
A.M.~Sirunyan, A.~Tumasyan
\vskip\cmsinstskip
\textbf{Institut f\"{u}r Hochenergiephysik, Wien, Austria}\\*[0pt]
W.~Adam, F.~Ambrogi, E.~Asilar, T.~Bergauer, J.~Brandstetter, E.~Brondolin, M.~Dragicevic, J.~Er\"{o}, A.~Escalante~Del~Valle, M.~Flechl, R.~Fr\"{u}hwirth\cmsAuthorMark{1}, V.M.~Ghete, J.~Hrubec, M.~Jeitler\cmsAuthorMark{1}, N.~Krammer, I.~Kr\"{a}tschmer, D.~Liko, T.~Madlener, I.~Mikulec, N.~Rad, H.~Rohringer, J.~Schieck\cmsAuthorMark{1}, R.~Sch\"{o}fbeck, M.~Spanring, D.~Spitzbart, A.~Taurok, W.~Waltenberger, J.~Wittmann, C.-E.~Wulz\cmsAuthorMark{1}, M.~Zarucki
% ...
\vskip\cmsinstskip
% ...
1:  Also at Vienna University of Technology, Vienna, Austria\\
% ...
\end{sloppypar}
\end{document}